\documentstyle[referee]{mn}
\input psfig.tex
\def\lsim{\lower.5ex\hbox{$\; \buildrel < \over \sim \;$}}
\def\gsim{\lower.5ex\hbox{$\; \buildrel > \over \sim \;$}}
\def \simeq{\lower.3ex\hbox{$\; \buildrel \sim \over - \;$}}
\def\ch{\lower-0.55ex\hbox{--}\kern-0.55em{\lower0.15ex\hbox{$h$}}}
\def\lh{\lower-0.55ex\hbox{--}\kern-0.55em{\lower0.15ex\hbox{$\lambda$}}}
\newif\ifAMStwofonts
\ifoldfss
  \ifCUPmtlplainloaded \else
    \NewTextAlphabet{textbfit} {cmbxti10} {}
    \NewTextAlphabet{textbfss} {cmssbx10} {}
    \NewMathAlphabet{mathbfit} {cmbxti10} {} 
    \NewMathAlphabet{mathbfss} {cmssbx10} {} 
  \fi
  \ifAMStwofonts
    \ifCUPmtlplainloaded \else
      \NewSymbolFont{upmath} {eurm10}
      \NewSymbolFont{AMSa} {msam10}
      \NewMathSymbol{\upi}     {0}{upmath}{19}
      \NewMathSymbol{\umu}     {0}{upmath}{16}
      \NewMathSymbol{\upartial}{0}{upmath}{40}
      \NewMathSymbol{\leqslant}{3}{AMSa}{36}
      \NewMathSymbol{\geqslant}{3}{AMSa}{3E}

      \let\leq=\leqslant 
       
    \fi
  \fi
\fi 
\ifnfssone
  \newmathalphabet{\mathit}
  \addtoversion{normal}{\mathit}{cmr}{m}{it}
  \addtoversion{bold}{\mathit}{cmr}{bx}{it}
  \newmathalphabet{\mathbfit} 
  \addtoversion{normal}{\mathbfit}{cmr}{bx}{it}
  \addtoversion{bold}{\mathbfit}{cmr}{bx}{it}
  \newmathalphabet{\mathbfss} 
  \addtoversion{normal}{\mathbfss}{cmss}{bx}{n}
  \addtoversion{bold}{\mathbfss}{cmss}{bx}{n}
  \ifAMStwofonts
    \ifCUPmtlplainloaded \else
      \UseAMStwoboldmath
      \makeatletter
      \new@mathgroup\upmath@group
      \define@mathgroup\mv@normal\upmath@group{eur}{m}{n}
      \define@mathgroup\mv@bold\upmath@group{eur}{b}{n}
      \edef\UPM{\hexnumber\upmath@group}
      \new@mathgroup\amsa@group
      \define@mathgroup\mv@normal\amsa@group{msa}{m}{n}
      \define@mathgroup\mv@bold\amsa@group{msa}{m}{n}
      \edef\AMSa{\hexnumber\amsa@group}
      \makeatother
      \mathchardef\upi="0\UPM19
      \mathchardef\umu="0\UPM16
      \mathchardef\upartial="0\UPM40
      \mathchardef\leqslant="3\AMSa36
      \mathchardef\geqslant="3\AMSa3E

      \let\leq=\leqslant 

    \fi
  \fi
\fi 

\ifnfsstwo
  \DeclareMathAlphabet{\mathbfit}{OT1}{cmr}{bx}{it}
  \SetMathAlphabet\mathbfit{bold}{OT1}{cmr}{bx}{it}
  \DeclareMathAlphabet{\mathbfss}{OT1}{cmss}{bx}{n}
  \SetMathAlphabet\mathbfss{bold}{OT1}{cmss}{bx}{n}
  \ifAMStwofonts
    \ifCUPmtlplainloaded \else
      \DeclareSymbolFont{UPM}{U}{eur}{m}{n}
      \SetSymbolFont{UPM}{bold}{U}{eur}{b}{n}
      \DeclareSymbolFont{AMSa}{U}{msa}{m}{n}
      \DeclareMathSymbol{\upi}{0}{UPM}{"19}
      \DeclareMathSymbol{\umu}{0}{UPM}{"16}
      \DeclareMathSymbol{\upartial}{0}{UPM}{"40}
      \DeclareMathSymbol{\leqslant}{3}{AMSa}{"36}
      \DeclareMathSymbol{\geqslant}{3}{AMSa}{"3E}

      \let\leq=\leqslant 

    \fi
  \fi
\fi 

\ifCUPmtlplainloaded \else
  \ifAMStwofonts \else 
    \def\upi{\pi}
    \def\umu{\mu}
    \def\upartial{\partial}
  \fi
\fi

\title{Scattering of Dirac Waves off Kerr  Black Holes}

\author[S. K. Chakrabarti and Banibrata Mukhopadhyay]
       {S. K. Chakrabarti and Banibrata Mukhopadhyay\\
S.N. Bose National Centre for Basic Sciences,\\
JD-Block, Sector III, Salt Lake, Calcutta 700091, India\\
chakraba@boson.bose.res.in and bm@boson.bose.res.in}

\date{Accepted for publication in MNRAS}
\begin{document}

\maketitle

\begin{abstract}

Chandrasekhar separated the Dirac equation for spinning and massive particles
in Kerr geometry into radial and angular parts. Here we solve the complete wave
equation and find out how the Dirac wave scatters off Kerr black holes.
The eigenfunctions, eigenvalues and reflection and transmission co-efficients
are computed. We compare the solutions with several parameters to
show how a spinning black hole distinguishes mass and energy of incoming waves.
Very close to the horizon the solutions become independent of the particle parameters
indicating an universality of the behaviour.

\end{abstract}


\section{Introduction}
 
Chandrasekhar (1976) separated Dirac equation in Kerr black hole geometry
into radial ($r$) and angular ($\theta$) parts. The radial
equations governing the radial wave-functions $R_{\pm \frac{1}{2}}$
corresponding to spin $\pm \frac{1}{2}$ are given by (with $\ch=1=G=c$):

$$
\Delta^{\frac{1}{2}}{\cal D}_{0} R_{- \frac{1}{2}}
= ( \lh + i m_p r) \Delta^{\frac{1}{2}} {R}_{+ \frac{1}{2}} ;\hskip0.7cm
\Delta^{\frac{1}{2}} {\cal D}_{0}^{\dag} \Delta^{1 \over 2}
{R}_{+{\frac{1}{2}}} = ( \lh - i m_p  r)  {R}_{-{1 \over 2}} ,
\eqno{(1)}
$$
where, the operators ${\cal D}_n$ and ${\cal D}_{n}^{\dag}$ are given by,
$$
{\cal D}_n = \partial_{r} + \frac {i K} {\Delta} + 2n \frac{(r-M)} {\Delta} ;\hskip0.7cm
{\cal D}_n^{\dag} = \partial_{r} - \frac {i K} {\Delta} + 2n \frac{(r-M)} {\Delta} ,
\eqno{(2)}
$$
and
$$
\Delta = r^2 + a^2 - 2Mr ;\hskip0.7cm
K=(r^2 + a^2)\sigma + am .
\eqno{(3)}
$$
Here, $a$ is the Kerr parameter, $n$ is an integer, $\sigma$ is the
frequency of incident wave, $M$ is the mass of the black hole, $m_p$ is the rest mass of
the Dirac particle, $\lh$ is the eigenvalue  which is the
separation constant of complete Dirac equation and $m$ is the azimuthal quantum
number.

The equations governing the angular wave-functions $S_{\pm \frac{1}{2}}$ corresponding
to spin $\pm \frac{1}{2}$ are given by:

$$
{\cal L}_{1 \over 2} S_{+ {1 \over 2}} = - (\lh -a m_p \cos \theta) S_{- {1 \over 2}}; \hskip0.7cm
{\cal L}_{1 \over 2}^{\dag} S_{- {1 \over 2}} = + (\lh +a m_p \cos \theta) S_{+
{1 \over 2}}
\eqno{(4)}
$$
where, the operators ${\cal L}_n$ and ${\cal L}_{n}^{\dag}$ are given by,
$$
{\cal L}_n = \partial_\theta + Q + n \cot \theta ;\hskip0.7cm
{\cal L}_n^{\dag} = \partial_\theta - Q + n \cot \theta
\eqno{(5)}
$$
and
$$
Q=a \sigma \sin \theta + m \  {\rm cosec} \  \theta .
\eqno{(6)}
$$
Combining eqs. 4, one obtains a second order angular eigenvalue equations which
admits exact solutions for spin-half particles when $\rho=\frac{m_p}{\sigma}=1$ (Chakrabarti, 1984),
$$
\lh^2 = (l+\frac{1}{2})^2 + a \sigma ( p+ 2m) + a^2 \sigma^2 \left [1-\frac{y^2}{2(l+1)
+a\sigma x} \right ] ,
\eqno{(7)}
$$
and
$$
{}_{1\over 2}S_{lm} =
{}_{1\over 2}Y_{lm} - \frac{a\sigma y}{2(l+1)+a\sigma x} {}_{1\over 2}Y_{l+1 m}
\eqno{(8)}
$$
where,
$$
p=F(l,l); \ \ \ x=F(l+1,l+1); \ \ \ y=F(l,l+1)
$$
and
$$
F(l_1,l_2)=[(2l_2+1)(2l_1+1)]^{\frac{1}{2}} <l_2 1 m 0|l_1 m>
[<l_2 1 \frac{1}{2} 0|l_1 \frac{1}{2}> +(-1)^{l_2-l}<l_2 1 m 0|l_1 m>
\eqno{(9)}
$$
$ \hskip8.0cm [<l_2 1 \frac{1}{2} 0|l_1 \frac{1}{2}> +(-1)^{l_2-l}
\rho \sqrt{2} <l_2 1 -\frac{1}{2} 1|l_1 \frac{1}{2}>] .  $\\

\medskip

\noindent Here, $<....|..>$ are the usual Clebsh-Gordon coefficients
and ${}_sY_{lm}$ are the standard spin-weighted spherical harmonics
(Chakrabarti, 1984; see also, Goldberg et al 1967, Breure et al, 1982) of spin 
$s$ and usual quantum numbers $l$ and $m$. When $\rho \ne \frac{m_p}{\sigma}=1$, one obtains the solutions
perturbatively with $a\sigma$ to be the perturbation parameter.
The detailed procedure to obtain eigenfunctions and eigenvalues
is in (Chakrabarti 1984) and is not described here.

The radial equations (1) are in coupled form. One can decouple them and express
the equation either in terms of spin up or spin down wave functions $R_{\pm \frac{1}{2}}$
but the expression loses its transparency. It is thus advisable to use the approach of
Chandrasekhar (1983) by changing the basis and independent variable $r$ to,
$$
r_{*} = r + \frac{2M r_+ + am/\sigma} {r_+ - r_-} {\rm log}
\left({r \over r_+} - 1\right) - \frac{2M r_- + am/\sigma} {r_+ - r_-} {\rm log} \left({r \over r_{-}} - 1\right)
\hskip0.2cm ( r > r_{+}).
\eqno{(10)}
$$
where,
$$
{d \over dr_{*}} = {\Delta \over \omega^{2}} {d \over dr}; \ \ \ \ \ \omega^2 =
r^2 + \alpha^2;\ \ \ \  \ \alpha^2 = a^2 + am/\sigma,
\eqno{(11)}
$$
to transform the set of coupled equations (eq. 1) into two independent one dimensional wave equations given by:
$$
\left({d \over dr_{*}} -  i \sigma\right)P_{+ {1 \over 2}} = \frac {\Delta^{1 \over 2}}
{\omega^2} (\lh - i m_p r) P_{- {1 \over 2}};\hskip0.7cm
\left({d \over dr_{*}} + i \sigma\right)P_{- {1 \over 2}} = \frac{\Delta^{1 \over 2} } {\omega^2}
(\lh + i m_p r) P_{+ {1 \over 2}} .
\eqno{(12)}
$$
Here, ${\cal D}_{0} = {\omega^{2} \over \Delta} ({d \over dr_*} + i \sigma)$
and ${\cal D}^{\dagger}_{0} ={\omega^2 \over \Delta} ({d \over dr_{*}} - i\sigma)$
were used and wave functions were redefined as $R_{- {1 \over 2}} = P_{- {1 \over 2}}$ and
$\Delta^{1 \over 2} R_{+ {1 \over 2}} = P_{+ {1 \over 2}}$.

\section{Solution Procedure}

We define a new variable,
$$
\theta = tan^{-1} (m_p r/{\lh}), \ \ \ \ {\rm which\ gives, } \ \ \ \
(\lh \pm i m_p r) = exp ({\pm i \theta}) \surd ({\lh}^{2} + m_p^2 r^2).
\eqno{(13)}
$$
Also define,
$$
P_{+ {1 \over 2}} = \psi_{+ {1 \over 2}}\  exp\left[-{1 \over 2} i\  tan^{-1} \left({{m_p r}
\over \lh}\right)\right];\hskip0.7cm
P_{- {1 \over 2}} = \psi_{- {1 \over 2}}\  exp\left[+{1 \over 2} i \ tan^{-1} \left({{m_p r}
\over \lh}\right)\right],
\eqno{(14)}
$$
and further choosing $\hat{r}_* = r_{*} + {1 \over 2\sigma} {\rm tan}^{-1}({m_p
r \over \lh})$ so that
$d{\hat r}_* = (1 + {\Delta \over \omega^2} {{\lh m_p} \over 2 \sigma} {1 \over
{\lh^2 + m_p^2 r^2}})dr_*$,
and $Z_\pm=\psi_{+\frac{1}{2}}\pm\psi_{-\frac{1}{2}}$
the above equations become,
$$
\left(\frac{d} {d{\hat r}_*} - W\right) Z_+ = i \sigma Z_- ;\hskip0.7cm
\left(\frac{d} {d{\hat r}_*} + W\right) Z_- = i \sigma Z_+  ,
\eqno{(15)}
$$
where,
$$
W = \frac{\Delta^{1 \over 2} (\lh^{2} + m_p^2 r^2)^{3/2} } { \omega^2 (\lh^2 + m_p^2 r^2)
+ \lh m_p \Delta/2\sigma}.
\eqno{(16)}
$$
>From these equations, we readily obtain a pair of independent one-dimensional wave equations,
$$
\left(\frac{d^2} {{d {\hat r}_*}^2} + \sigma^2\right) Z_\pm = V_\pm Z_\pm ; \ \ \ \ {\rm where} \ \ \ \
V_{\pm} = W^{2} \pm {dW \over d\hat{r}_{*}}.
\eqno{(17)}
$$
By transformation of the variable from $r$ to $r_{*}$ (and ${\hat{r}}_{*}$)
the horizon is shifted from $r=r_+$ to ${\hat {r}}_{*}= -\infty$ unless
$\sigma \leq \sigma_s=-am/2Mr_+$ (eq. 10). In this connection, it is customary to define
$\sigma_c$ where $\alpha^2=0$ (eq. 11). Thus, $\sigma_c=-m/a$. If $\sigma \leq \sigma_s$,
super-radiation is expected for particles with integral spins but not for those
with half-integral spin (Chandrasekhar 1983). Thus, we concentrate on the region where, $\sigma>\sigma_s$.

The choice of parameters is generally made in such a way that there is
a significant interaction between the particle and the black hole, i.e., when the Compton
wavelength of the incoming wave is of the same order as the outer horizon of
the Kerr black hole. Similarly, the frequency of the incoming
particle (or wave) should be of the same order as inverse of light
crossing time of the radius of the black hole. These yield,
$$
m_p \sim \sigma \sim [M + \sqrt(M^2 - a^2)]^{-1}.
\eqno{(18)}
$$
Thus, we need to deal with quantum black holes to get `interesting' results.
There are two cases of interest: (1) the waves do not `hit' the potential barrier
and (2) the waves do hit the potential barrier.  First, we replace the
potential barrier by a large number of steps as in the step-barrier problem in
quantum mechanics. Fig. 1 shows one such example of the potential barrier
$V_+$ (eq. 17) which is drawn for $a=0.5$, $m_p=0.8$ and $\sigma=0.8$. In reality
we use tens of thousands of steps with suitable variable widths so that the
steps become indistinguishable from the actual function.
The solution of (17) at $n$th step can be written as (Davydov 1976),
$$
Z_{+,n}  = A_{n} exp [ik_n {\hat r}_{*, n}] + B_{n} exp [-ik_n{\hat r}_{*, n}]
\eqno{(19)}
$$
when energy of the wave is greater than the height of the potential barrier.
The standard junction condition is given as (Davydov 1976),
$$
Z_{+, n}=Z_{+, n+1} \ \ \ \ {\rm and} \ \ \ \
\frac{dZ_{+}}{d{\hat r}_*}|_n = \frac{dZ_{+}}{d{\hat r}_*}|_{n+1} .
\eqno{(20)}
$$
\begin {figure}
\vbox{
\vskip 10.0cm
\hskip 0.0cm
\centerline{
\psfig{figure=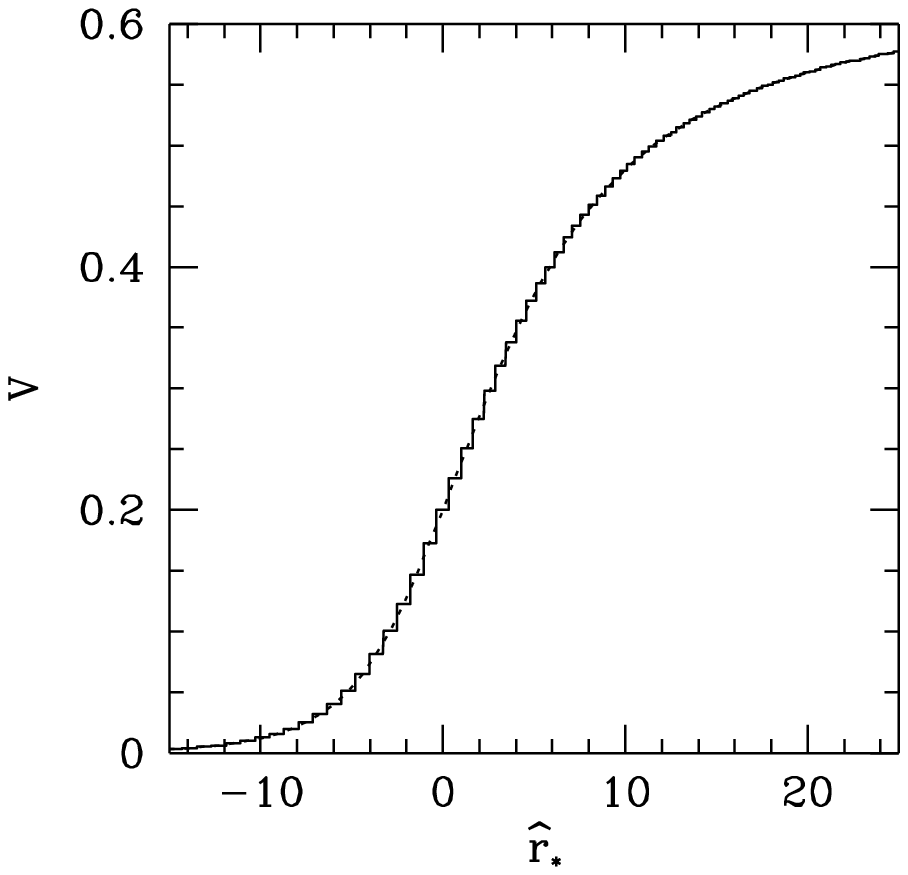,height=10truecm,width=10truecm}}}
\vspace{-1.5cm}
\noindent {\small {\bf Fig. 1} : 
Behaviour of $V_+$ (smooth solid curve) for $a=0.5,\ m_p=0.8,\ \sigma=0.8$.
This is approximated as a collection of steps. In realily tens of thousand steps were used
with varying step size which mimic the potential with arbitrary accuracy. }
\end{figure}

The reflection and transmission co-efficients at $n$th junction are given by:
$$
R_n=  \frac{A_{n+1} (k_{n+1}-k_n)+B_{n+1}(k_{n+1}+k_n)}{A_{n+1}(k_{n+1}+k_n)+B_{n+1}(k_{n+1}-k_n)}; \ \ T_n=1-R_n
\eqno{(21)}
$$
At each of the $n$ steps these conditions were used to connect solutions at successive steps.
Here, $k$ is the wave number ($k=\sqrt{\sigma^2-V_\pm}$) of the wave and $k_n$ is
its value at $n$th step. We use the  `no-reflection' inner boundary condition:
$R \rightarrow 0$ at ${\hat r}_* \rightarrow -\infty $.

For the cases where waves hit on the potential barrier, inside the barrier
(where $\sigma^2 < V_+$) we use the wave function of the form
$$
Z_{+,n}  = A_{n} exp [-\alpha_n {\hat r}_{*, n}] + B_{n} exp [\alpha_n{\hat r}_{*, n}]
\eqno{(22)}
$$
where, $\alpha_n=\sqrt{V_\pm-\sigma^2}$, as in usual quantum mechanics.

\begin {figure}
\vbox{
\vskip 10.0cm
\hskip 0.0cm
\centerline{
\psfig{figure=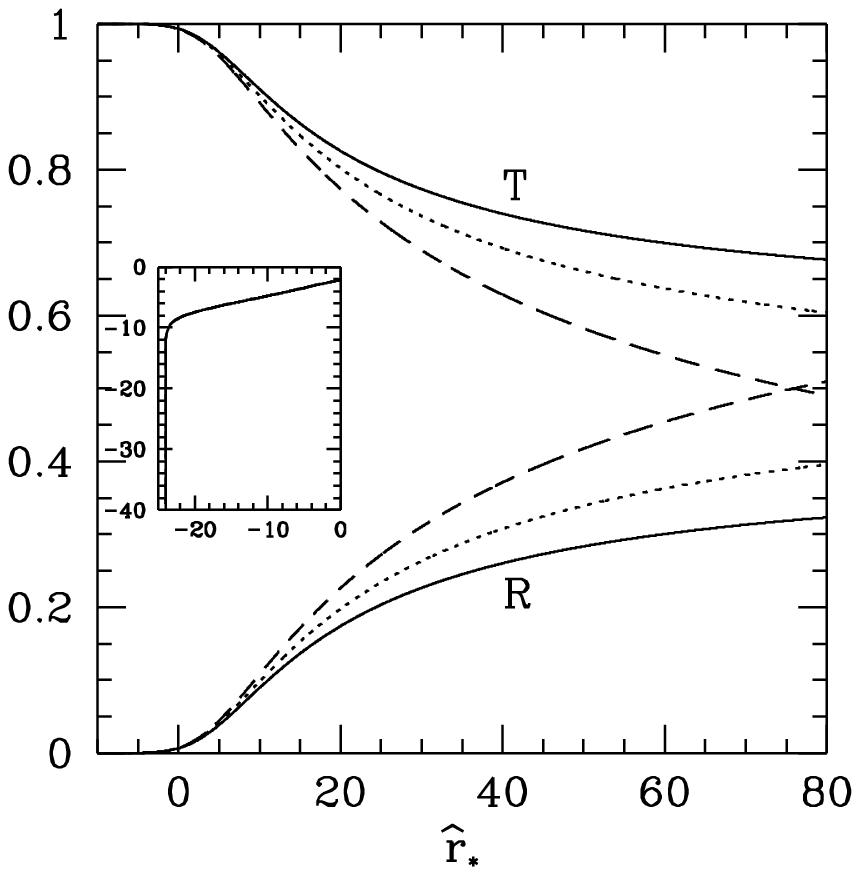,height=10truecm,width=10truecm}}}
\vspace{-1.5cm}
\end{figure}
\begin{figure}
\noindent {\small {\bf Fig. 2a}:
Reflection ($R$) and transmission ($T$) coefficients of waves
with varying mass as functions of ${\hat r}_*$. $m_p=0.78$ (solid),
$m_p=0.79$ (dotted) and $m_p=0.80$ (long-dashed) are used.
Other parameters are $a=0.5$ and $\sigma=0.8$. Inset shows
$R$ in logarithmic scale which falls off exponentially just outside the horizon.}
\end{figure}

\begin {figure}
\vbox{
\vskip 10.0cm
\hskip 0.0cm
\centerline{
\psfig{figure=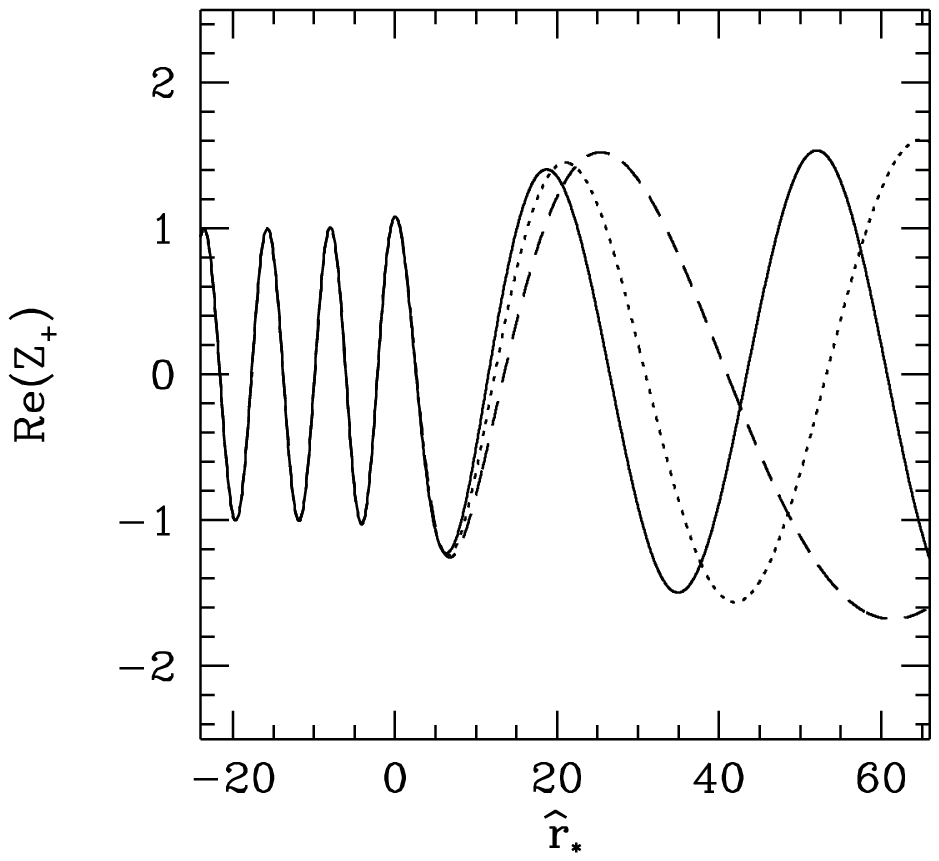,height=10truecm,width=10truecm}}}
\vspace{-1.5cm}
\end{figure}

\begin{figure}
\noindent {\small {\bf Fig. 2b}:
Amplitude of Re($Z_+$) of waves
with varying mass as functions of ${\hat r}_*$. $m_p=0.78$ (solid),
$m_p=0.79$ (dotted) and $m_p=0.80$ (long-dashed) are used.
Other parameters are $a=0.5$ and $\sigma=0.8$. }
\end{figure}

\section{Examples of Solutions}

Fig. 2a shows three solutions [amplitudes of Re($Z_+$)] for parameters: $a=0.5$, $\sigma=0.8$ and
$m_p=0.78,\ 0.79, $ and $0.80$ respectively in solid, dotted and long-dashed curves.
The energy $\sigma^2$ is always higher compared to the height of the potential barrier (Fig. 1)
and therefore the particles do not `hit' the barrier. $k$ goes up and therefore the wavelength
goes down monotonically as the wave approaches a black hole. It is to be noted that
though ours is apparantly a `crude' method, it has flexibility and is capable of presenting
insight into the problem, surpassing any other method such as ODE solver packages.
This is because one can choose (a) variable steps depending on steepness of the potential to
ensure uniform accuracy, and at the same time (b) virtually infinite number of steps to follow the
potential as closely as possible. For instance, in the inset, we show $R$ in logarithmic scale
very close to the horizon. All the three curves marge, indicating that the
solutions are independent of the mass of the particle and a closer inspection shows that here, the slope
of the curve depends only on $\sigma$. The exponential dependence of $R_n$ close to the horizon
becomes obvious. Asymptotically, $V_\pm=m_p^2$
(eq. 17), thus, as $m_p$ goes down, the wavelength goes down. In Fig. 2b, we present the
instantaneous values of the reflection $R$ and transmission $T$ coefficients (i.e., $R_n$ and $T_n$ of eq. 21)
for the same three cases. As the particle mass is decreased, $k$ goes up and  corresponding $R$
goes down consistent with the limit that as $k \rightarrow \infty$, there would be no reflection
at all as in a quantum mechanical problem.

\begin {figure}
\vbox{
\vskip 10.0cm
\hskip 0.0cm
\centerline{
\psfig{figure=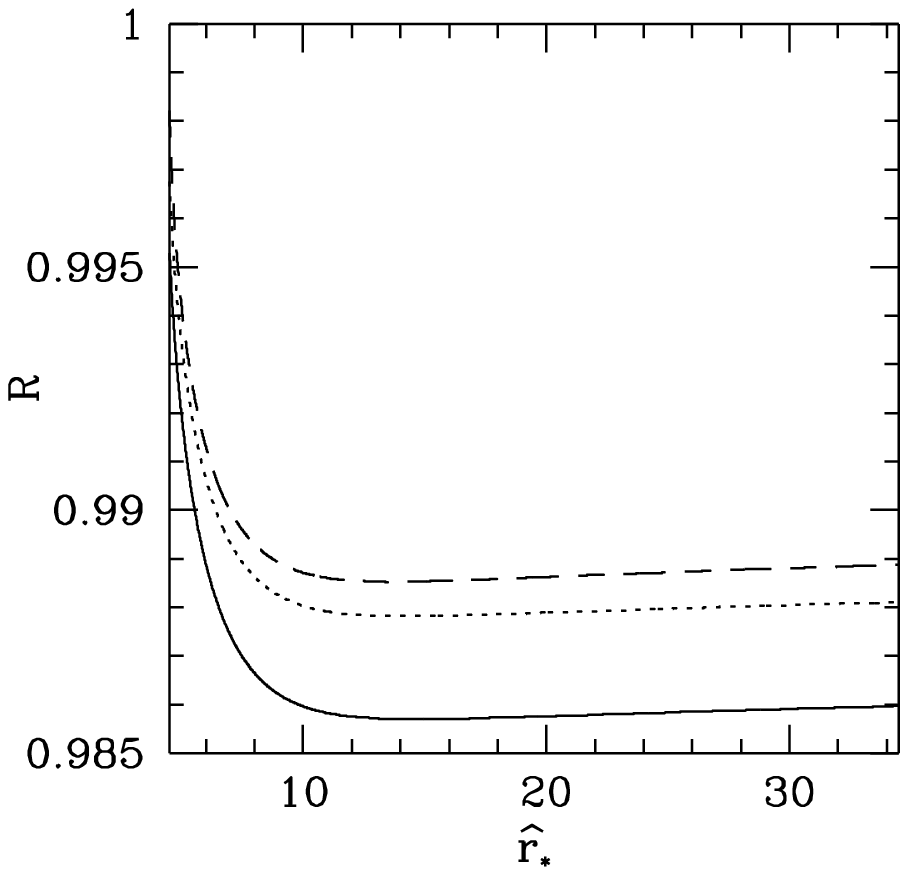,height=10truecm,width=10truecm}}}
\vspace{-1.5cm}
\end{figure}
\begin {figure}
\noindent {\small {\bf Fig. 3a}}: Reflection ($R$) coefficient of waves
with varying mass as functions of ${\hat r}_*$. $m_p=0.16$ (solid), $m_p=0.164$ (dotted)
and $m_p=0.168$ (long-dashed) are used. 
Other parameters are $a=0.95$ and $\sigma=0.168$.
\end{figure}

\begin {figure}
\vbox{
\vskip 10.0cm
\hskip 0.0cm
\centerline{
\psfig{figure=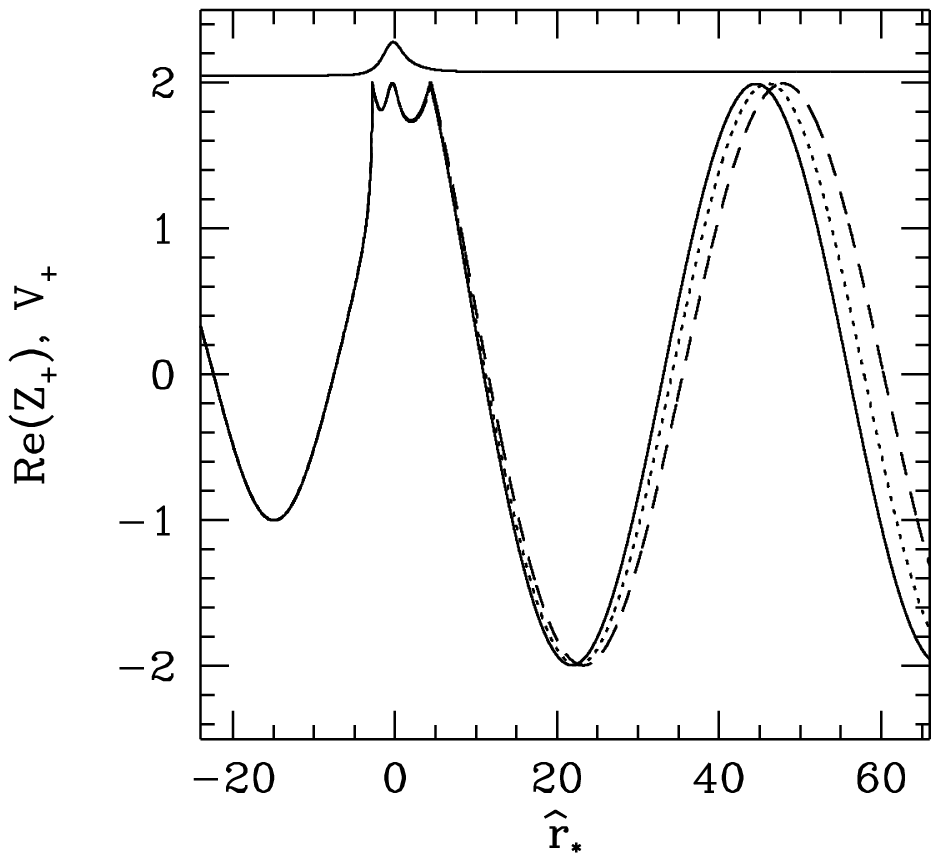,height=10truecm,width=10truecm}}}
\vspace{-1.5cm}
\end{figure}

\begin {figure}
\noindent {\small {\bf Fig. 3b}}: Amplitude of Re($Z_+$) of waves
with varying mass as functions of ${\hat r}_*$. $m_p=0.16$ (solid), $m_p=0.164$ (dotted)
and $m_p=0.168$ (long-dashed) are used. Nature of potential with $m_p=0.168$ is drawn shifting
vertically by 2.05 unit for clarity. Other parameters are $a=0.95$ and $\sigma=0.168$.
\end{figure}

Figs. 3(a-b) compare a few solutions where the incoming particles
`hit' the potential barrier. We choose, $a=0.95$, $\sigma=0.168$ and mass of the particle
$m_p=0.16,\ 0.164, \ 0.168$ respectively in solid, dotted and long-dashed curves.
Inside the barrier, the wave decays before coming back to a sinusoidal behaviour,
before entering into a black hole. In Fig. 3b, we plotted the potential (shifted by
2.05 along vertical axis for clarity).  Here too, the reflection coefficient
goes down as $k$ goes up consistent with the classical result that as the
barrier height goes up more and more, reflection is taking place strongly. Note however,
that the reflection is close to a hundred percent. Tunneling causes only a
few percent to be lost into the black hole.

\begin {figure}
\vbox{
\vskip 6.0cm
\hskip 0.0cm
\centerline{
\psfig{figure=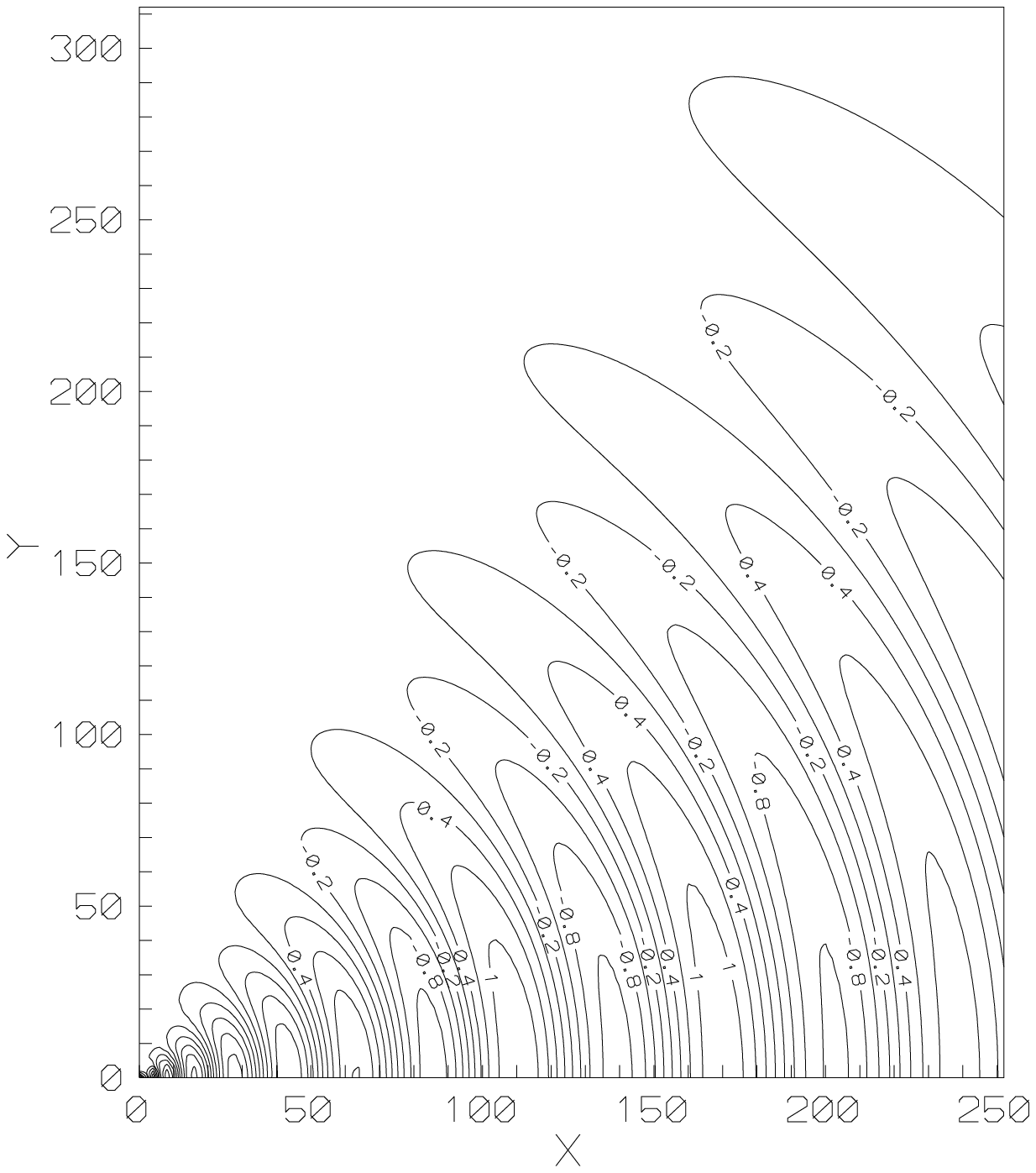,height=14truecm,width=14truecm}}}
\vspace{-1.5cm}
\end{figure}
\begin{figure}
\noindent {\small {\bf Fig. 4a}: Contours of constant amplitude 
are plotted in the meridional plane around a black hole. Radial direction on equatorial plane
is  along $X$ axis and the vertical direction and along $Y$.
Both radial and theta solutions have been combined. Parameters
are $a=0.5$, $m_p=0.8$ and $\sigma=0.8$.}
\end{figure}

\begin {figure}
\vbox{
\vskip 0.0cm
\hskip 0.0cm
\centerline{
\psfig{figure=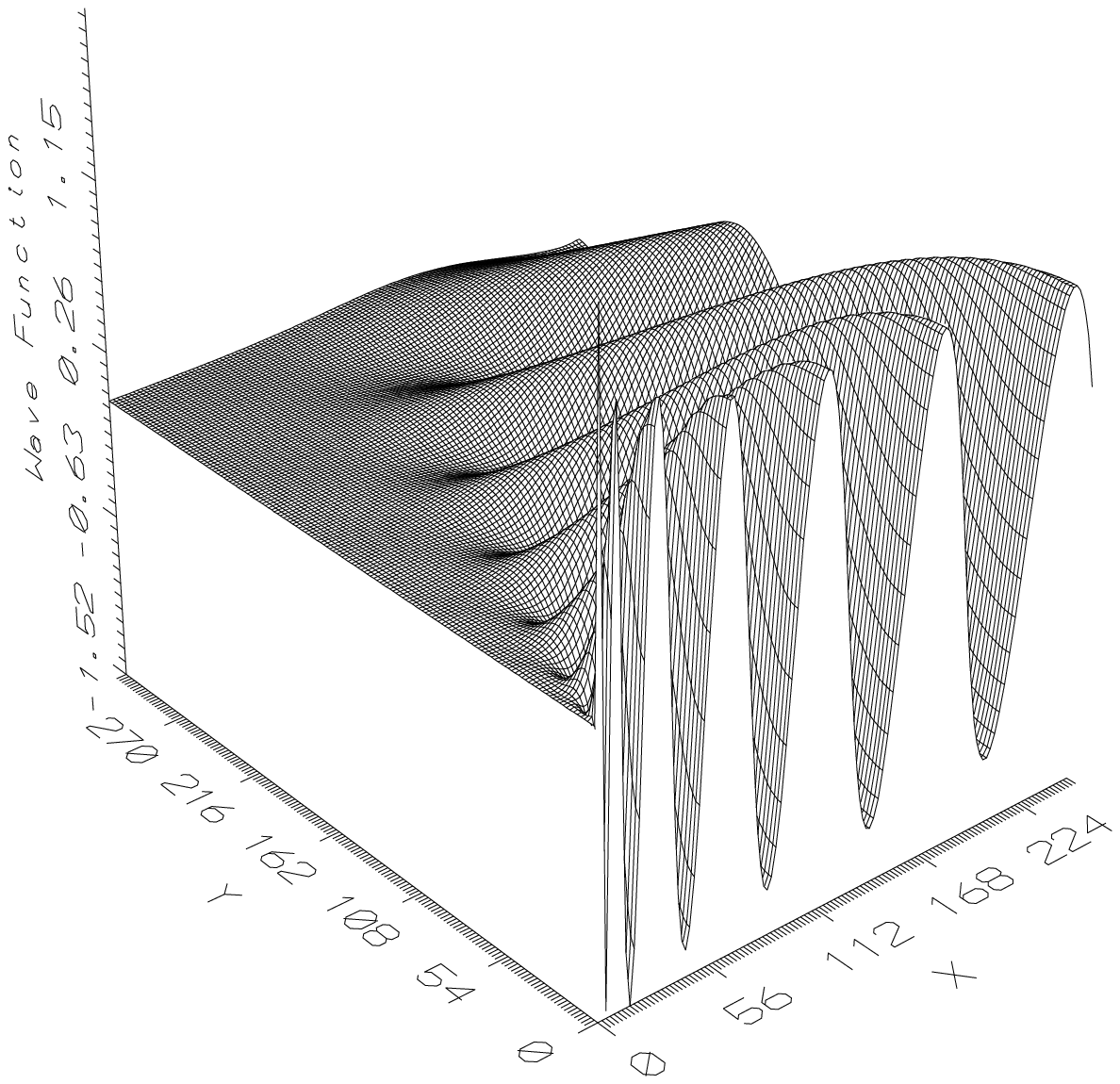,height=14truecm,width=14truecm}}}
\vspace{-1.5cm}
\end{figure}
\begin{figure}
\noindent {\small {\bf Fig. 4b}: Three dimensional view of $R_{-1/2}S_{-1/2}$ are plotted
in the meridional plane around a black hole. 
Both radial and theta solutions have been combined. Parameters
are $a=0.5$, $m_p=0.8$ and $\sigma=0.8$.}
\end{figure}

Figs. 4(a-b) show the nature of the complete wave function
when both the radial and the angular solutions (Chakrabarti 1984) are included.
Fig. 4a shows contours of constant amplitude of the
wave ($R_{-1/2} S_{-1/2}$) in the meridional plane -- $X$ is along
radial direction in the equatorial plane  and $Y$ is along the vertical direction.
The parameters are $a=0.5$, $m_p=0.8$ and $\sigma=0.8$.
Some levels are marked. Two successive contours have amplitude
difference of $0.1$. In Fig. 4b a three-dimensional nature of
the complete solution is given. Both of these figures clearly
show how the wavelength varies with distance. Amplitude of the spherical
wave coming from a large distance also gets weaker along the
vertical axis and the wave is forced to fall generally along the equatorial
plane, possibly due to the dragging of the inertial frame.

\section{Conclusion}

Scatterings of massive, spin-half particles from a
spinning black hole has been studied with particular emphasis to the
nature of the radial wave functions and the reflection and
transmission coefficients. Well known quantum mechanical step-potential approach is
used but we applied it successfully to a complex problem of barrier penetration
in a spacetime around a spinning black hole. Among
significant observations, we find that the wave function and $R$, and $T$
behave similarly close to the horizon independent of the initial
parameter, such as the particle mass $m_p$. Particles of different
mass scatter off to a large distance completely differently, thus giving an impression that
a black hole could be treated as a mass spectrograph! When the energy
of the particle becomes higher compared to the rest mass, the reflection
coefficient diminishes as it should it. Similar to a barrier
penetration problem, the reflection coefficient becomes close to a
hundred percent when the wave hits the potential barrier.
Another significant observation is that the reflection and transmission
coefficients are functions of the radial coordinates. This is
understood easily because of the very nature of the potential barrier
which is strongly space dependent which we approximate as a collection
of steps. Combining with the solution of theta-equation, we
find that the wave-amplitude vanishes close to the vertical axis,
possibly due to the frame-dragging effects.


{}

\end{document}